\documentclass[%
superscriptaddress,
 amsmath,amssymb,
 aps,
prb,
twocolumn,
]{revtex4-2}
\usepackage[italicdiff]{physics}
\usepackage[dvipdfmx]{graphicx}
\usepackage{dcolumn,amsmath,txfonts,here,subfigure}
\usepackage{bm}
\usepackage{color}

\newcommand{\rA}{{\mathrm{A}}}

\newcommand{\rF}{{\mathrm{F}}}

\newcommand{\rL}{{\mathrm{L}}}

\newcommand{\rR}{{\mathrm{R}}}
\newcommand{\rS}{{\mathrm{S}}}
\newcommand{\rT}{{\mathrm{T}}}
\newcommand{\rU}{{\mathrm{U}}}

\newcommand{\rX}{{\mathrm{X}}}
\newcommand{\rY}{{\mathrm{Y}}}

\newcommand{\sgn}{\mathrm{sgn}}

\begin{document}

\preprint{APS/123-QED}

\title{Coherent electron splitting in interacting chiral edge channels}


\author{Eiki Iyoda}%
\affiliation{
Department of Physics, Tokai University, \\
4-1-1, Kitakaname, Hiratsuka-shi, Kanagawa 259-1292, Japan
}
\author{Takase Shimizu}%
\author{Masayuki Hashisaka}%
\affiliation{
NTT Basic Research Laboratories, NTT Corporation, \\
3-1 Morinosato-Wakamiya, Atsugi, Kanagawa 243-0198, Japan
}
\affiliation{
Institute for Solid State Physics, The University of Tokyo, \\
5-1-5 Kashiwanoha, Kashiwa, Chiba 277-8581, Japan
}

\date{\today}

\begin{abstract}
This paper theoretically studies the quantum coherence of an electronic state on an artificial chiral Tomonaga-Luttinger (TL) liquid. Coulomb interaction between copropagating integer quantum Hall edge channels causes the TL liquid nature of charge excitations, resulting in the splitting of an electronic state into bosonic eigenmodes. We investigate the single-electron coherence under the splitting process by calculating the Aharonov-Bohm (AB) oscillations in an electronic Mach-Zehnder interferometer employing copropagating spin-up and spin-down edge channels as the interference paths. We investigate the voltage bias dependence of the AB oscillations at zero temperature, taking the inter-channel interaction into account using the bosonization technique. The calculation results of the visibility and the phase of the AB oscillations show non-monotonical bias dependence when the copropagating channels are electrostatically asymmetric. These observations are interpreted as the signatures of the second-order interference between the fractionalized spin excitations with different phase evolutions. We also report finite entanglement entropy between the bosonic eigenmodes split from an electron, which presents an analogy between the `electron splitting' in a TL liquid and the Cooper-pair splitting at a superconducting junction.
\end{abstract}

\pacs{Valid PACS appear here}
\maketitle


\section{Introduction}

When an electron is impinged on a quantum many-body system hosting non-trivial charge and spin carriers, it often results in generating multiple carriers~\cite{Andreev1963, Tinkham1996, Hashisaka2021, Cohen2023}. The most representative example is the Andreev reflection at a superconductor
junction, where an electron injection results in a hole reflection to form a Cooper pair in the superconductor. Besides, a Tomonaga-Luttinger (TL) liquid hosts bosonic elementary excitations (eigenmodes) stemming from the pronounced electron correlation in a one-dimensional (1D) electron system~\cite{Tomonaga1950,Luttinger1963,Haldane1981,Giamarchi2003,Chang2003}. The TL liquid nature gives rise to the splitting of an electron, e.g., the spin-charge separation and the charge fractionalization~\cite{Auslaender2002,Jompol2009,Berg2009,Bocquillon2013,Kamata2014,Inoue2014,Freulon2015,Hashisaka2017,Hashisaka2018}, providing another opportunity to examine the intriguing electron dynamics at a boundary of a quantum many-body system.

Here, we examine the coherent electron splitting into multiple bosonic eigenmodes on an artificial chiral TL liquid of copropagating spin-up and spin-down edge channels in the integer quantum Hall (QH) state~\cite{Berg2009,Bocquillon2013,Inoue2014,Freulon2015,Hashisaka2017,Hashisaka2018}. Recent experiments demonstrated the Aharonov-Bohm (AB) interference in an electronic Mach-Zehnder interferometer (MZI) employing copropagating spin-up and spin-down edge channels as the interference paths~\cite{Nakajima2013,Shimizu2020,Shimizu2023,Shimizu2024Joint}. In the single-particle picture, the observed AB interference can be interpreted as the spin precession during the propagation along the spin-full 1D channel. In an actual device, on the other hand, the inter-channel interaction induces `electron splitting', breaking the interpretation of the single-particle picture. Our motivation is to examine the impact of the electron splitting process on the coherent spin precession. We analyze the bias dependence of the AB oscillations considering the inter-channel interaction using the bosonization technique. The calculation results of the visibility and the phase of the AB oscillations vary non-monotonically as a function of the bias when the copropagating channels are electrostatically asymmetric. These observations reflect the second-order interference between the fractionalized spin excitations propagating at different speeds. Our theoretical model, taking only the dominant Coulomb interaction between the interference paths into account, explains the major features of the experimental observations reported in the companion paper ~\cite{Shimizu2024Joint}. 

\begin{figure}
	\centering
	\includegraphics[width=0.95\linewidth]{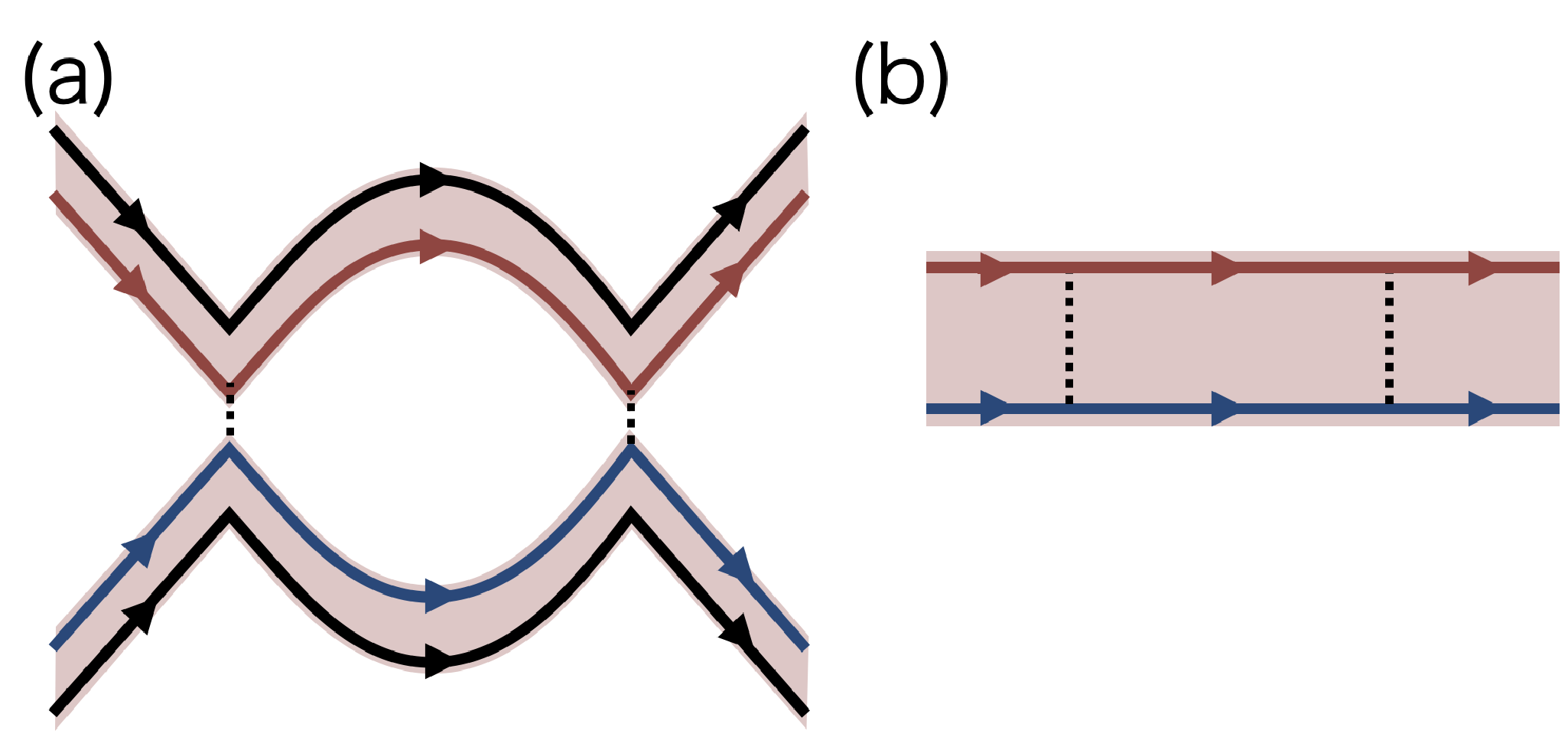}
	\caption{
	Schematic of the MZI. Red and blue lines represent the interference paths of the upper and lower channels, respectively. Black lines are the environmental channels copropagating with the interference path. The red region indicates the presence of Coulomb interaction.
	(a) Conventional MZI.
	(b) Our setup.
	}
	\label{fig:MZI_setup}
\end{figure}

It is worth noting that the MZI under our study contrasts with the conventional MZIs~\cite{Ji2003,Neder2006,Roulleau2007,Litvin2008,Bieri2009,Sukhorukov2007,Chalker2007,Neder2008,Youn2008,Helzel2015,Jo2022}. While the inter-channel interaction presents between the interference paths in our setup [Fig.~\ref{fig:MZI_setup}(b)], electron dynamics in conventional MZIs [Fig.~\ref{fig:MZI_setup}(a)] is often dominated by the interaction between the paths and the environmental channels. Thus, our study sheds light on the electron coherence through the splitting process, which is different from previous studies addressing decoherence due to inter-channel interaction~\cite{Levkivskyi2008}. Coherent splitting of a single information carrier can generate quantum entanglement between the resultant multiple carriers, as in the case of the Cooper-pair splitting at a superconducting junction~\cite{Hofstetter2009}. We estimate finite entanglement entropy between fractionalized spins and point out an analogy between the present electron splitting process and the Cooper-pair splitting.

This paper is structured as follows. Section II formulates the mode transformation using the bosonization technique and presents the eigenmodes in interacting edge channels. We also calculate the entanglement entropy between the eigenmodes generated from the initial electronic state. Section III analyzes the AB interference in the MZI composed of the interacting channels to examine the impact of the splitting process on electron coherence. Section IV discusses the comparison of the calculations with the experimental results reported in the companion paper. Section V is the summary of this paper. In addition, the Appendix explains the details of the calculations omitted in the main text.
\section{Coherent spin splitting}
\label{sec:fractionalize}

Inter-channel interaction couples charge excitations in copropagating QH edge channels, transforming the excitation eigenmodes in the system. In this section, we formulate the mode transformation using the bosonization technique and discuss electron splitting on the artificial chiral TL liquid.

\subsection{Setup: Copropagating chiral channels}
\label{subsec:setup}

We consider copropagating upper ($j=\rU$, spin up) and lower ($j=\rL$, spin down) chiral edge channels. The Hamiltonian of the system is given by $\mathcal{H}:=\mathcal{H}_0+\mathcal{H}_\mathrm{int}$, where $\mathcal{H}_0$ and $\mathcal{H}_\mathrm{int}$ represent the kinetic energy and the Coulomb interaction, respectively:  
\begin{align}
\mathcal{H}_0
&:=
-iv_\rF\sum_{j=\rU,\rL}
\int dx \psi_j^\dag(x) \partial_x \psi_j(x),
\\
\mathcal{H}_\mathrm{int}
&:=
\frac{1}{2}
\sum_{i,j}
\int\int dx dy
U_{ij}(x-y)\rho_i(x)\rho_j(x).
\end{align}
Here, $\psi_j(x)$ is the annihilation operator for an electron, $v_\mathrm{F}$ is the Fermi velocity,  $\rho_j(x)$ is the electron density at $x$ on channel $j$, and $U_{ij}(x-y)$ is the Coulomb interaction between charge densities $\rho_i(x)$ and $\rho_j(y)$.

The electron annihilation operator can be rewritten in the bosonized form ~\cite{Delft1998,Giamarchi2003} as
\begin{align}
\label{eq:bosonize1}
\psi_i&\sim F_i e^{i\phi_i},
\\
\label{eq:bosonize2}
\phi_i&:=
\varphi_i+2\pi p_i x+\tilde{\phi}_i,
\\
\label{eq:bosonize3}
\tilde{\phi}_i
&:=
\sum_{k>0}\sqrt{\frac{2\pi}{Wk}}[a_i(k)e^{ikx}+a_i^\dag e^{-ikx}],
\end{align}
where $F_i$ ($i=\rU,\rL$) is the Klein factor~\cite{Delft1998} for the upper/lower channel, $p_i$ is the annihilation operator of the zero mode in channel $i$, $\varphi_i$ is its conjugate operator, $a_i(k)$ is the annihilation operator of the boson field of wavenumber $k$, and $W$ is the size of the system (we will take the thermodynamic limit $W\rightarrow\infty$ later). These operators satisfy the following commutation relations:
\begin{align}
[\phi_i(x),\phi_i(y)]
&=
i\pi \sgn(x-y),
\\
[p_i,\varphi_i]&=\frac{i}{W},
\\
[a_i(k),a_j(k^\prime)]=[a^\dag_i(k),a^\dag_j(k^\prime)]&=0,
\\
[a_i(k),a^\dag_j(k^\prime)]&=\delta_{kk^\prime}\delta_{ij}.
\end{align}
The Hamiltonian $\mathcal{H}$ can be described with the boson fields as~\cite{Levkivskyi2008}
\begin{align}
\label{eq:hamiltonian}
\mathcal{H}
&=
\frac{1}{2\pi}
\sum_{i,j=\rU,\rL}
\sum_k
k
V_{ij}(k)a_i^\dag(k) a_j(k)
+
\frac{W}{2}\sum_{i,j=\rU,\rL}V_{ij}(0)p_{i}p_j.
\end{align}

Neglecting the dispersion of the interaction as $V_{ij}(k)=V_{ij}(0)$, we obtain the ground state of $\mathcal{H}$ as the simultaneous eigenstate of $\mathcal{H}$ and $p_i$:
\begin{align}
\mathcal{H}|0\rangle&=E_0|0\rangle,
\\
p_i|0\rangle&=Q_i|0\rangle,
\end{align}
where $Q_i$ is the vacuum charge. Since the bosonic excitations are absent, the ground state energy is expressed only by the zero modes $Q_i$ as
\begin{align}
E_0=\frac{W}{2}\sum_{ij}V_{ij}(0)Q_i Q_j,
\end{align}
and the chemical potential as
\begin{align}
\Delta\mu_i=
\frac{1}{W}\frac{\delta E_0}{\delta Q_i}
=
\sum_j V_{ij}(0)Q_j.
\end{align}

\subsection{Fast and slow modes}
\label{subsec:eigenmodes}

Coulomb interaction in the system can be described in the matrix form as
\begin{align}
\label{eq:interaction_matrix}
V
:=
\left[
 \begin{matrix}
   V_{\mathrm{U} \mathrm{U}} & V_{\mathrm{U} \mathrm{L}} \\
   V_{\mathrm{L} \mathrm{U}} & V_{\mathrm{L} \mathrm{L}} 
 \end{matrix}
\right].
\end{align}
In general, the interactions $V_\mathrm{UU}$ and $V_\mathrm{LL}$ differ ($V_\mathrm{UU}\neq V_\mathrm{LL}$). When we use the interaction matrix $\tilde{V}=V/(2\pi)$, the 1D wave equation of charge densities is given by 
\begin{align}
\frac{\partial}{\partial t}
\begin{pmatrix}
\rho_\uparrow \\ 
\rho_\downarrow \\
\end{pmatrix}
=-\tilde{V}\frac{\partial}{\partial x}
\begin{pmatrix}
\rho_\uparrow \\ 
\rho_\downarrow \\
\end{pmatrix}.
\end{align}
We obtain the eigenmodes of the present system by diagonalizing $V$ with the unitary matrix $S$ as
\begin{align}
\label{eq:interaction_2ch1}
V
&=
S\Lambda S^\dag,
\\
\label{eq:interaction_2ch2}
\Lambda&:=
2\pi
\left[
 \begin{matrix}
   u & 0 \\
   0 & v 
 \end{matrix}
\right],
\end{align}
where $u$ and $v$ are the group velocities of the eigenmodes. When the inter-channel interaction $V_\mathrm{UL} = V_\mathrm{LU}$ is comparable with or larger than the intra-channel interactions, we find $u\gg v$. Thus, we call the eigenmode with $u$ the `fast (plasmon) mode' and that with $v$ the `slow (dipole) mode.'
The emergence of the fast/slow modes has been well studied both experimentally and theoretically~\cite{Berg2009,Levkivskyi2008,Hashisaka2018}. 

We diagonalize the Hamiltonian [Eq.~(\ref{eq:hamiltonian})] by replacing the boson field $a_i(k)$ with the annihilartion operator of the eigenmode $b_j(k)$ that relates to $a_i(k)$ as
\begin{align}
a_i(k)
&=
\sum_{j=u,v} S_{ij}b_j(k)~~~(i=\mathrm{U,L}).
\end{align}
Substituting this into Eq.~(\ref{eq:bosonize3}) leads to
\begin{align}
\label{eq:bosonize_uv}
\tilde{\phi_i}
&=
\phi_{iu}+\phi_{iv},
\\
\phi_{iu}
&:=
\sum_{k>0}
\sqrt{\frac{2\pi}{Wk}}
\left[
S_{iu}b_u(k)e^{ikx}
+
S_{iu}^*b_u^\dag(k)e^{-ikx}
\right],
\\
\phi_{iv}
&:=
\sum_{k>0}
\sqrt{\frac{2\pi}{Wk}}
\left[
S_{iv}b_v(k)e^{ikx}
+
S_{iv}^*b_v^\dag(k)e^{-ikx}
\right].
\end{align}

Since $V$ is real and symmetric, we can rewrite $V$ with parameters $a,b,c\in\mathbb{R}$ as follows:
\begin{align}
\label{eq:specialV2}
V&=
aI+bZ+cX
=
\left[
 \begin{matrix}
   a+b & c \\
   c & a-b 
 \end{matrix}
\right],
\\
\label{eq:specialV_S2}
S&=
\left[
 \begin{matrix}
   \cos\frac{\theta}{2} & \sin\frac{\theta}{2} \\
   \sin\frac{\theta}{2} & -\cos\frac{\theta}{2} 
 \end{matrix}
\right],
\\
\label{eq:specialV_S2cos}
\cos\theta
&:=
\frac{b}{\sqrt{b^2+c^2}},
\\
\label{eq:specialV_S2sin}
\sin\theta
&:=
\frac{c}{\sqrt{b^2+c^2}},
\end{align}
where $I$, $Z$, and $X$ are the identity matrix, the Pauli $Z$ matrix, and the Pauli $X$ matrix, respectively.
The parameter $b$ represents the difference between the upper and the lower channels.
We also define the following quantities:
\begin{align}
\label{eq:delta_def}
\delta_i&:=1-S_{\rU i}(S_{\rL i})^*-S_{\rL i}(S_{\rU i})^*.
\end{align}
The unitarity of the matrix $S$ gives $\delta_u+\delta_v=2$.
Here, $\delta_u=1-\sin\theta$ characterizes the asymmetry between the upper and the lower channels in the way that $\delta_u= 0$ at $b = 0$ and $\delta_u$ monotonically increases with $b$. Therefore, we refer to $\delta_u$ as the `asymmetry parameter'.
In general, $b$, and hence $\delta_u$ is small but finite for the integer QH edge channels in a GaAs/AlGaAs heterostructure~\cite{Hashisaka2018}.

If we consider the completely symmetric interaction ($\delta_u= 0$) as the simplest example, we find
\begin{align}
\label{eq:specialV}
V&=
\pi
\left[
 \begin{matrix}
   u+v & u-v \\
   u-v & u+v 
 \end{matrix}
\right],
\\
\label{eq:specialV_S}
S&=
\frac{1}{\sqrt{2}}
\left[
 \begin{matrix}
   1 & 1 \\
   1 & -1 
 \end{matrix}
\right].
\end{align}
The eigenmodes are represented as
\begin{align}
\label{eq:eigen_b_fast}
b_u(k)
&=
\frac{1}{\sqrt{2}}
(a_\rU(k)+a_\rL(k)),
\\
\label{eq:eigen_b_slow}
b_v(k)
&=
\frac{1}{\sqrt{2}}
(a_\rU(k)-a_\rL(k)).
\end{align}
In this case, $b_u$ corresponds to the pure charge mode and $b_v$ the pure spin mode.

\subsection{Injection of single electron}
\label{subsec:injection_se}

\begin{figure}
	\centering
	\includegraphics[width=1\linewidth]{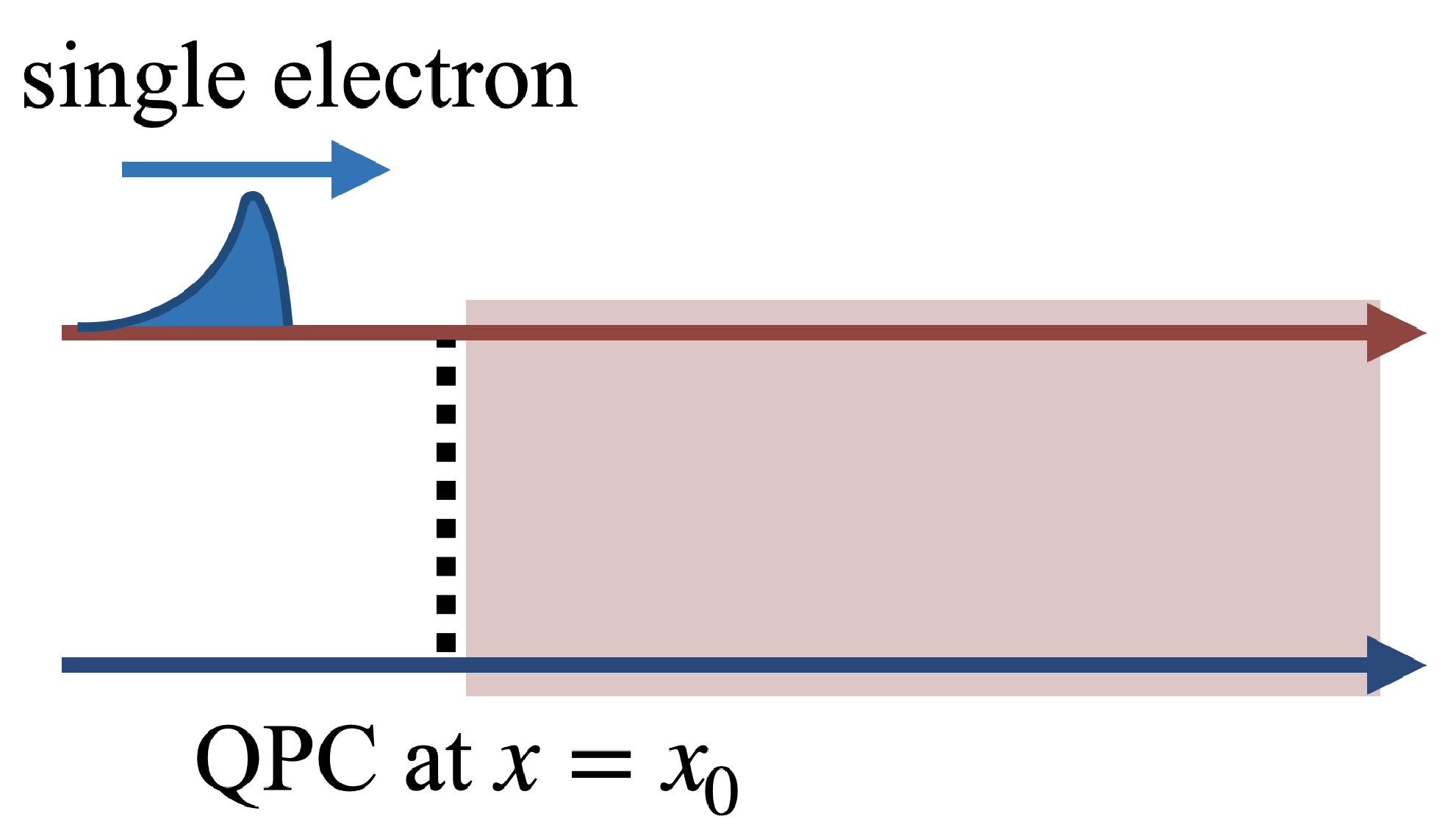}
	\caption{
	A single electron impinged on a QPC at $x=x_0$.
	}
	\label{fig:single_electron_injection}
\end{figure}

In this subsection, we consider a quantum point contact (QPC) prepared between the copropagating channels, as shown in Fig.~\ref{fig:single_electron_injection}. We assume that the inter-channel interaction is `switched on' downstream of the QPC and neglect the impact of the interaction on the tunneling probability, for simplicity~\cite{Rebora2020,Acciai2022}. We describe the scattering matrix of the QPC as 
\begin{align}
\label{eq:scattering_matrix}
\left[
 \begin{matrix}
 \psi_\rU(x_0 +\varepsilon) \\
  \psi_\rL(x_0 +\varepsilon)
 \end{matrix}
\right]
=
\left[
 \begin{matrix}
 \sqrt{T} & i\sqrt{R} \\
 i\sqrt{R} & \sqrt{T}
 \end{matrix}
\right]
\left[
 \begin{matrix}
 \psi_\rU(x_0 -\varepsilon) \\
  \psi_\rL(x_0 -\varepsilon)
 \end{matrix}
\right],
\end{align}
where $x_0$ is the QPC position,
$\varepsilon$ is an infinitesimal small positive,
and $T,R\in\mathbb{R}$ are the transmission and reflection probabilities, respectively. In this formalism, an electron scattering event at the QPC creates a superposition of the electron states existing on the upper spin-up and the lower spin-down channels.

\begin{figure*}
	\centering
	\includegraphics[width=1\linewidth]{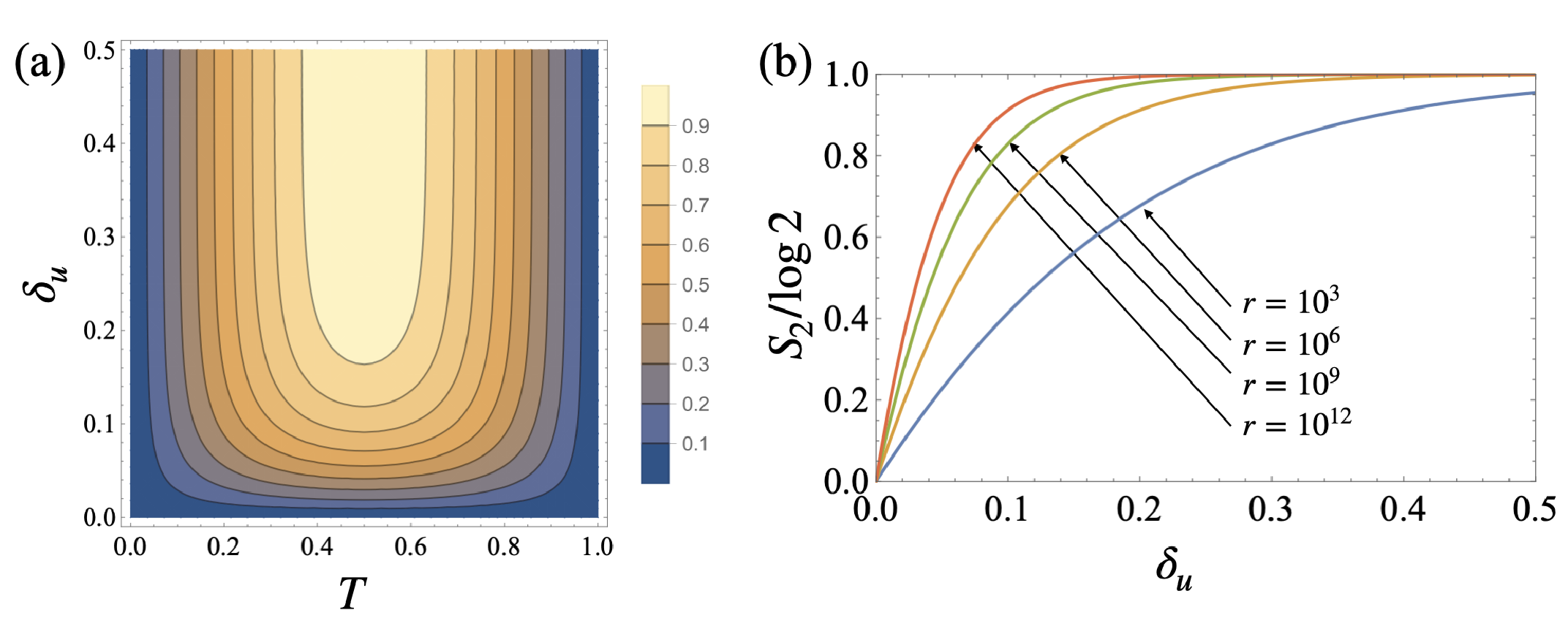}
	\caption{
	The $\delta_u$ and $T$ dependence of the Renyi-2 EE $S_2/\log 2$.
	(a) Contour plot of $S_2/\log 2$ ($r=10^7$).
	(b) The $\delta_u$ dependence of $S_2/\log 2$ ($T=1/2$ and $r=10^3,10^6,10^9,10^{12}$).
	}
	\label{fig:EE}
\end{figure*}

Let $t=0$ be the time immediately after the scattering event.
The state at $t=0$ is given by
\begin{align}
\label{eq:QPC_Smat_initial_state}
|\psi(0)\rangle
=
\left(
\sqrt{T}\psi_\rU^\dag(x_0+\varepsilon)
+
i\sqrt{R}\psi_\rL^\dag(x_0+\varepsilon)
\right)
|0\rangle,
\end{align}
and that at $t~(>0)$ is
\begin{align}
|\psi(t)\rangle
&=
e^{-iHt/\hbar}
\left(
\sqrt{T}\psi_\rU^\dag(x_0+\varepsilon)
+
i\sqrt{R}\psi_\rL^\dag(x_0+\varepsilon)
\right)
|0\rangle
\\
\label{eq:single_electron_psit}
&=
e^{-iE_0t/\hbar}
\left(
\sqrt{T}\psi_\rU^\dag(x_0+\varepsilon,t)
+
i\sqrt{R}\psi_\rL^\dag(x_0+\varepsilon,t)
\right)
|0\rangle,
\end{align}
where $\psi_i(x,t)$ is the Heisenberg representation of $\psi_i(x)$.

From Eqs.~(\ref{eq:bosonize1}) and (\ref{eq:bosonize_uv}), we find
\begin{align}
\psi_i^\dag(x_0+\varepsilon,t)|0\rangle 
\sim
\exp[i\phi_{i u}(x_0+\varepsilon,t)+i\phi_{i v}(x_0+\varepsilon,t)]|0\rangle,
\end{align}
for an electron on the upper ($i = U$) or the lower ($i = L$) channel. Therefore, both $\psi_\rU^\dag(x_0+\varepsilon,t)|0\rangle $ and $\psi_\rL^\dag(x_0+\varepsilon,t)|0\rangle $ in Eq.~(\ref{eq:single_electron_psit}) split into the fast/slow modes, providing us an opportunity to study the evolution of the single-electron superposition state under fractionalization. 

\subsection{Entanglement entropy}
\label{subsec:entanglement}

The coherent splitting of an electronic state in the copropagating channels [Eq. (\ref{eq:single_electron_psit})] reminds us of the Cooper-pair splitting at a superconductor junction. When a Cooper pair comes out from a superconductor to split into two normal leads, the resultant state is a spin singlet of spatially separated electrons. This Cooper-pair splitting has been intensively studied as a way of preparing quantum entanglement in condensed matter. In semiconductor systems, electron scattering can also generate electron-hole entanglement~\cite{Beenakker2006,Zhang2024}. In terms of possible analogies with these systems, it is worth examining quantum entanglement between the eigenmodes in the artificial chiral TL liquid.

Quantum entanglement yields a finite von Neumann entanglement entropy (EE) $S_{\mathrm{vN}}$ between two excitations. Here, we calculate the Renyi-2 EE between the fast and slow modes split from a superposition state:
\begin{align}
\label{eq:Renyi-2EE}
S_2=-\log\left(
T^2+R^2+2TR 
\langle e^{i\phi_{\rU u}}e^{-i\phi_{\rL u}}\rangle
\langle e^{i\phi_{\rL u}}e^{-i\phi_{\rU u}}\rangle
\right),
\end{align}
which gives a lower bound of $S_{\mathrm{vN}}$ (see Appendix~\ref{sec:appendixEE}.) The correlation function in the right hand side is calculated as
\begin{align}
\langle e^{i\phi_{\rU u}}e^{-i\phi_{\rL u}}\rangle
\simeq
r^{-\delta_u/2},
\end{align}
where we define $r$ as
\begin{align}
\frac{2\pi}{W}\sum_{k>0} \frac{1}{k}\simeq
\int_{k_\mathrm{min}}^{k_\mathrm{max}} \frac{dk}{k}
=
\log\frac{k_\mathrm{max}}{k_\mathrm{min}}
=:
\log r.
\end{align}
Here, $k_\mathrm{max}$ is the high energy cutoff, which effectively represents the band width, 
and $k_\mathrm{min}$ is of the order of $2\pi/W$.
The ratio $r=k_\mathrm{max}/k_\mathrm{min}$ diverges in the thermodynamic limit~($W\rightarrow \infty$), resulting in
\begin{align}
r^{-\delta_u}\rightarrow
\begin{cases}
1 ~~~(\delta_u=0)\\
0 ~~~(\delta_u>0)
\end{cases}.
\end{align}
We rewrite Eq. (\ref{eq:Renyi-2EE}) as
\begin{align}
\label{eq:RenyiEE}
S_2=-\log
\left[
1-2T(1-T)(1-r^{-\delta_u})
\right],
\end{align}
where we used $T+R=1$. Note that $S_2$ is independent of time $t$ since the fast and slow modes are the (independent) eigenmodes in the copropagating channels.

Figure~\ref{fig:EE}(a) is a colored contour plot of $S_2/\log 2$ as a function of $\delta_u$ and $T$. Finite EE is observed at $0<T<1$ and $\delta_u>0$. Since the splitting of a superposition state generates the entanglement, we find $S_2=0$ at $T=0$ and $T=1$. At $0<T<1$, $S_2$ increases with $\delta_u$ in the range of $0\leq \delta_u \leq 1/2$, as indicated in Fig.~\ref{fig:EE}(b) for several $r$ values at $T=1/2$. The monotonical increase of $S_2$ from $S_2 = 0$ indicates that the entanglement relates to the spin degree of freedom carried by the eigenmodes. As represented by Eq.~(\ref{eq:eigen_b_fast}), at $\delta_u=0$, the fast mode carries no spin information [in other words, no information about on which (upper or lower) channel an electron exists]. This situation corresponds to the complete spin-charge separation in the artificial TL liquid. When $\delta_u$ increases from $\delta_u=0$, the spin degree of freedom carried by the fast mode becomes finite, causing quantum entanglement between the fractionalized spin excitations. Thus, quantum entanglement results from the electrostatic asymmetry between the copropagating channels.

\section{Mach-Zehnder Interferometer}
\label{sec:MZI}

Copropagating edge channels with tunnel couplings at two positions form an MZI, as illustrated in Fig. \ref{fig:MZI_setup}(b). This setup enables us to investigate the coherent evolution of the spin-up and spin-down superposition state prepared at the entrance QPC [see Eq.~(\ref{eq:QPC_Smat_initial_state})] by observing the Aharonov-Bohm (AB) interference.

\subsection{Setup: MZI}
\label{subsec:setup_MZI}

We consider copropagating spin-up and spin-down chiral QH edge channels with Coulomb interaction represented by Eq.~(\ref{eq:interaction_matrix}). Two QPCs at $x=x_\rL$ and $x_\rR$ induce tunnel couplings between the channels to form an MZI, as shown in Fig.~\ref{fig:MZI_setup}(b). Below, we calculate the electric current flowing from the upper to the lower channels through the interferometer. The magnetic fluxes penetrating the aperture between the channels induce the AB oscillations of the electric current. 

Note the major difference between our MZI and the conventional MZI~[Fig.~\ref{fig:MZI_setup}(a)] is the presence of the Coulomb interaction between the interference paths. Except for this point, our formulation is similar to the conventional MZI~\cite{Levkivskyi2008,Levkivskyi2010}.

The tunneling Hamiltonian $\mathcal{H}_\rT$ is defined as 
\begin{align}
\mathcal{H}_\rT
&:=
\hat{A}+\hat{A}^\dag=
\sum_{l=\rL,\rR}
(\hat{A}_l+\hat{A}_l^\dag),
\\
\label{eq:Al}
\hat{A}_l
&:=
t_l\psi^\dag_\rL(x_l)\psi_\rU(x_l),
\end{align}
where $t_l$ ($l=\rL,\rR$) is the tunneling amplitude at $x=x_l$. Here, we set $x_\rL=0$ and $x_\rR=L~(>0)$ for simplicity. Tunneling probability is given by $\Gamma_l:=|t_l|^2$. We introduce the AB phase $\varphi$ as the phase of the tunneling terms as
\begin{align}
\label{eq:ABphase1}
t_l t_{l^\prime}^*
&=
|t_l t_{l^\prime}|
e^{i\varphi_{ll^\prime}},
\\
\label{eq:ABphase2}
\varphi_{\rL\rR}=-\varphi_{\rR\rL}&=\varphi,
\\
\label{eq:ABphase3}
\varphi_{\rR\rR}=\varphi_{\rL\rL}&=0.
\end{align}

\subsection{Aharonov-Bohm oscillation visibility}
\label{subsec:def_current}

The tunneling current operator is obtained by summing up the number of particles scattered from the upper to the lower channels as 
\begin{align}
\hat{I}
&:=i(\hat{A}^\dag-\hat{A}),
\\
\hat{A}
&:=
\hat{A}_\mathrm{L}+\hat{A}_\mathrm{R}.
\end{align}
We calculate the current through the MZI within the framework of the non-equilibrium Green's function, taking into account the lowest-order contribution of the tunneling term $\mathcal{H}_\rT$. The average current can be expressed by
\begin{align}
\label{eq:average_current_sum}
I
&:=
\sum_{ll^\prime}
I_{ll^\prime},
\\
\label{eq:current_lowest_T}
I_{ll^\prime}
&:=
\int dt \langle [A^\dag_l(t), A_{l^\prime}(0)]\rangle,
\end{align}
where the ground state average is defined as $\langle \cdots\rangle:=\langle 0|\cdots|0\rangle$.
We also define $I^>_{ll^\prime}$ and $I^<_{ll^\prime}$ for convenience as
\begin{align}
I^>_{ll^\prime}
&:=
\int dt \langle A^\dag_l(t)A_{l^\prime}(0)\rangle,
\\
I^<_{ll^\prime}
&:=
\int dt \langle A_{l^\prime}(0)A^\dag_l(t)\rangle,
\end{align}
which lead to $I=
\sum_{ll^\prime}
(I^>_{ll^\prime}-I^<_{ll^\prime})$.

The differential conductance is described as
\begin{align}
\label{eq:def_differential}
\mathcal{G}&:=\frac{dI}{d\Delta\mu},
\end{align}
where $\Delta\mu:=\Delta \mu_\rU-\Delta\mu_\rL$.
We define the visibility of the AB oscillations of $\mathcal{G}$ as
\begin{align}
\label{eq:def_visibility}
V_\mathcal{G}&:=\frac{
\mathcal{G}_\mathrm{max}-\mathcal{G}_\mathrm{min}
}{\mathcal{G}_\mathrm{max}+\mathcal{G}_\mathrm{min}}
,
\end{align}
where $\mathcal{G}_\mathrm{max}$($ \mathcal{G}_\mathrm{min}$) is the maximum(minimum) value of $\mathcal{G}$. We will evaluate the electron coherence in the present setup by calculating the visibility and the phase of the AB oscillations. 

\subsection{Calculations for the average currnet}
\label{subsec:calcG:Current}

We evaluate the average electric current through the MZI by calculating the following integral:
\begin{align}
\label{eq:current_lesser}
I_{ll^\prime}^<
=
|t_lt_{l^\prime}|e^{-i\varphi_{ll^\prime}}
\int dt 
\langle 
\psi_\rU^\dag(x_l,t)
\psi_\rL(x_l,t)
\psi_\rL^\dag(x_{l^\prime},0)
\psi_\rU(x_{l^\prime},0)
\rangle.
\end{align}

In the conventional MZI case, the four-point function in Eq.~(\ref{eq:current_lesser}) can be factorized due to the absence of the interaction between the paths as
\begin{align}
\nonumber
&\langle 
\psi_\rU^\dag(x_l,t)
\psi_\rL(x_l,t)
\psi_\rL^\dag(x_{l^\prime},0)
\psi_\rU(x_{l^\prime},0)
\rangle
\\
=&
\langle 
\psi_\rU^\dag(x_l,t)
\psi_\rU(x_{l^\prime},0)
\rangle
\langle 
\psi_\rL(x_l,t)
\psi_\rL^\dag(x_{l^\prime},0)
\rangle.
\end{align}
In contrast, our setup with the inter-channel interaction requires us to calculate the four-point function directly. Letting $X=(x_l,t)$ and $X^\prime=(x_{l^\prime},0)$, the annihilation operators for electrons on the upper and lower channels are written as
\begin{align}
\label{eq:psitU}
\psi_\rU(X)
&\sim
e^{-i\Delta \mu_\rU t+2\pi iQ_\rU x_l} e^{i{\phi}_{\rU u}(X)+i{\phi}_{\rU v}(X)},
\\
\label{eq:psitL}
\psi_\rL(X)
&\sim
e^{-i\Delta \mu_\rL t+2\pi iQ_\rL x_l} e^{i{\phi_{\rL u}}(X)+i{\phi_{\rL v}}(X)},
\end{align}
where the Klein factor is omitted for simplicity. Substituting Eqs.~(\ref{eq:psitU}) and (\ref{eq:psitL}) into the four-point function, we obtain
\begin{align}
\label{eq:com_without_phase}
\prod_{j=u,v}
\langle
e^{-i\phi_{\rU j}}
e^{i\phi_{\rL j}}
e^{-i\phi^\prime_{\rL j}}
e^{i\phi^\prime_{\rU j}}
\rangle,
\end{align}
where 
$\phi_{ij}$ ($\phi_{ij}^\prime$) denotes $\phi_{ij}(X)$ ($\phi_{ij}(X^\prime)$).
We here omit the phase factors such as $e^{-i\Delta \mu_\rU t+2\pi iQ_\rU x_l} $, for which we will discuss later. Using the Baker-Campbell-Hausdorff formula repeatedly, we convert Eq.~(\ref{eq:com_without_phase}) to 
\begin{align}
\label{eq:com_without_phase2}
\prod_{j=u,v}
\left\{
\langle
e^{-i\phi_{\rU j}+i\phi_{\rL j}-i\phi^\prime_{\rL j}+i\phi^\prime_{\rU j}}
\rangle
e^{\frac{1}{2}[\phi_{\rU j},\phi_{\rL j}]}
e^{\frac{1}{2}[\phi^\prime_{\rL j},\phi^\prime_{\rU j}]}
e^{\frac{1}{2}[\phi_{\rU j}-\phi_{\rL j}, \phi^\prime_{\rU j}-\phi^\prime_{\rL j}]}
\right\}.
\end{align}
This equation can be written as
\begin{align}
\label{eq:result2ch4pt}
\frac{\mathcal{F}_{ll^\prime}(\delta_u,t)}{
(X_u-X_u^\prime)^{\delta_u}
(X_v-X_v^\prime)^{\delta_v}
},
\end{align}
where $\mathcal{F}_{ll^\prime}(\delta,t)$ is given by
\begin{align}
\label{eq:phase_factor_F}
\mathcal{F}_{ll^\prime}(\delta,t)&=
(-1)\times
\begin{cases}
\exp(\pi i \delta)& l\neq l^\prime\land \frac{L}{u}<t<\frac{L}{v}\\
1&\mathrm{otherwise}.
\end{cases}
\end{align}
Here, $X_u$ and $X_v$ are defined as
$X_u:=
x-ut$, $X_v
:=
x-vt$. For details of the calculations in this paragraph, see Appendix~\ref{sec:appendix2ch}.

The phase factors in Eqs.~(\ref{eq:psitU})(\ref{eq:psitL}) can be simplified as
$e^{i\Delta \mu (t-\Delta t_{ll^\prime})}$, where
$\Delta t_{ll^\prime}$ is defined as
\begin{align}
\Delta t_{ll^\prime}
&:=
\frac{x_l-x_{l^\prime}}{c},
\\
\label{eq:cinv1}
c^{-1}
&:=
u^{-1}
\left(
|S_{\rU u}|^2-S_{\rL u} S^*_{\rU u}
\right)
+
v^{-1}
\left(
|S_{\rU v}|^2-S_{\rL v} S^*_{\rU v}
\right).
\end{align}
The effective velocity $c$ is obtained as
\begin{align}
c^{-1}
&=
v^{-1}
\left(
1-\left(1-\frac{v}{u}\right)
\frac{1+\cos\theta-\sin\theta}{2}
\right),
\end{align}
by substituting Eq.~(\ref{eq:specialV_S2}) to Eq.~(\ref{eq:cinv1}).
We note that $c\simeq v$ when $\delta_u$ is small, 
namely, $\theta\simeq\pi/2$, $\cos\theta\simeq 0$, and $\sin\theta\simeq 1$.

From the above calculations, 
Eq.~(\ref{eq:current_lesser}) becomes 
\begin{align}
\label{eq:current_lesser2}
I^<_{ll^\prime}
&=
\frac{
|t_lt_{l^\prime}|
}{4\pi^2}
e^{-i\varphi_{ll^\prime}}
\int dt 
\mathcal{F}_{ll^\prime}(\delta,t)
\frac{e^{i\Delta \mu (t-\Delta t_{ll^\prime})}}{
(X_u-X_u^\prime)^{\delta_u}
(X_v-X_v^\prime)^{\delta_v}
}.
\end{align}

\subsection{Analytical results}
\label{subsec:calG:analytic}

\subsubsection{Symmetric channels}

When the copropagating channels are symmetric electrostatically, i.e., $\delta_u=0$ and $\delta_v=2$, we can calculate the average current straightforwardly as 
\begin{align}
\label{eq:current_symmetricV}
I
&=
\frac{
\Gamma_\rL+\Gamma_\rR
+2
\sqrt{\Gamma_\rL\Gamma_\rR}
\cos(\varphi)
}{\pi v^2} \Delta \mu,
\end{align}
which is linear with respect to $\Delta \mu$.
The visibility $V_\mathcal{G}$ [Eq.~(\ref{eq:def_visibility})] is given by
\begin{align}
\label{eq:visibility_symmetric}
V_\mathcal{G}=\frac{2
\sqrt{\Gamma_\rL\Gamma_\rR}}{\Gamma_\rL+\Gamma_\rR},
\end{align}
which does not depend on $\Delta\mu$ and equals $1$ when $\Gamma_\rL=\Gamma_\rR$.
Equation~(\ref{eq:visibility_symmetric}) is the same as the visibility for noninteracting electrons, meaning that Coulomb interaction does not affect the interference pattern when the channels are symmetric. This result contrasts with the conventional MZI case, where the lobe pattern appears even when the inter-channel interaction with environmental copropagating channels is symmetric.

\subsubsection{Asymmetric channels}

We next consider the asymmetric channels with $\delta_u\neq 0$. Since $\delta_u$ is small within experimentally reasonable settings, as stated in Sec.~\ref{subsec:eigenmodes}, we perform the following calculations assuming $\delta_u \ll 1$.

The differential conductance $\mathcal{G}$ is obtained by summing up the contributions of the diagonal and the off-diagonal components of the tunneling current. The diagonal components' contribution is described as [see Eqs.~(\ref{eq:average_current_sum}),~(\ref{eq:current_lowest_T}),~(\ref{eq:def_differential}),~(\ref{eq:current_lesser2})] 
\begin{align}
\frac{dI_{ll}}{d\Delta\mu}
=
\frac{\Gamma_l}{\pi u^{\delta_u} v^{\delta_v}},
\end{align}
which does not depend on $\Delta \mu$. 
On the other hand, that of the off-diagonal components at $\delta_u \ll 1$ and $u\gg v$ is given by
\begin{align}
\nonumber
&\frac{d(I_{\rR\rL}+I_{\rL\rR})}{d\Delta\mu}
\\
=&
\frac{
\sqrt{\Gamma_\rL\Gamma_\rR}
}{\pi^2u^{\delta_u} v^{2-{\delta_u}}}
\mathrm{Im}\left[
\int_C d\tau
\mathcal{F}_{\rR\rL}(\delta,t)
\frac{
\cos
\left
(\overline{\Delta\mu}
\left(
\tau -1
\right)
-\varphi
\right)
}{
\left(
-\tau +i\eta
\right)^{\delta_u}
\left(
1-\tau+i\eta
\right)^{1-{\delta_u}}
}
\right],
\label{eq:current_2ch_ILR}
\end{align}
where $\tau:=vt/L$ is the normalized time, $\overline{\Delta\mu}:=L\Delta\mu/v$ is the normalized bias, and $\eta$ is an infinitesimal small positive representing the causality. We note that the branch cut extends from $\tau=i\eta$  to $\tau=1+i\eta$ in the above integral, and the contour of integration goes counterclockwise around the branch cut. Because the numerator of the integrand contains $\overline{\Delta\mu}$, the AB oscillations reflect the bias voltage.

Simplifying the above integration, we obtain 
\begin{align}
\mathcal{G}&:=
\frac{\Gamma_\rL+\Gamma_\rR}{\pi u^{\delta_u} v^{\delta_v}}
\left(1-\frac{2\sqrt{\Gamma_\rL\Gamma_\rR}}{\Gamma_{\rL}+\Gamma_\rR} \tilde{I}(\varphi)\right),
\\
\label{eq:visibility_Itilde}
\tilde{I}(\varphi)
&:=
\frac{\cos(\pi\delta_u)\sin(\pi\delta_u)}{\pi}
(
I_C
\cos\varphi
+
I_S
\sin\varphi),
\\
\label{eq:visibility_IC}
I_C
&:=
\int_0^1
dt
t^{-\delta_u}(1-t)^{\delta_u-1}
\cos(\overline{\Delta\mu}(t-1)),
\\
\label{eq:visibility_IS}
I_S
&:=
\int_0^1
dt
t^{-\delta_u}(1-t)^{\delta_u-1}
\sin(\overline{\Delta\mu}(t-1)).
\end{align}
Then, the visibility is written as
\begin{align}
\label{eq:visibility1}
V_{\mathcal{G}}
=
\frac{2\sqrt{\Gamma_\rL\Gamma_\rR}}{\Gamma_{\rL}+\Gamma_\rR} 
\frac{\cos(\pi\delta_u)\sin(\pi\delta_u)}{\pi}
\sqrt{I_C^2+I_S^2}.
\end{align}
Non-trivial effects due to the Coulomb interaction are encapsulated within 
Eqs.~(\ref{eq:visibility_Itilde}),(\ref{eq:visibility_IC}), and (\ref{eq:visibility_IS}).
We note that the phase $\varphi_0$ of the AB oscillations is represented as $\varphi_0:=\mathrm{arg}(I_C+iI_S)$.

The integrals in $I_C$ and $I_S$ can be performed analytically 
and the result is represented by using the generalized hypergeometric function $_2F_3$ as
\begin{align}
I_C
=&
\frac{\pi}{\sin({\delta_u}\pi)}
\left[
\cos(\overline{\Delta\mu})
F_C
+
(1-{\delta_u})\overline{\Delta\mu}
\sin(\overline{\Delta\mu})
F_S
\right],
\\
I_S
=&
\frac{\pi}{\sin({\delta_u}\pi)}
\left[
-\sin(\overline{\Delta\mu})
F_C
+
(1-{\delta_u})\overline{\Delta\mu}
\cos(\overline{\Delta\mu})
F_S
\right],
\\
F_C:=&
~_2 F_3\left(\frac{1-{\delta_u}}{2},\frac{2-{\delta_u}}{2}; \frac{1}{2},\frac{1}{2},1; -\frac{{\overline{\Delta\mu}}^2}{4}\right),
\\
F_S:=&
~_2 F_3\left(\frac{2-{\delta_u}}{2},\frac{3-{\delta_u}}{2}; 1,\frac{3}{2},\frac{3}{2}; -\frac{{\overline{\Delta\mu}}^2}{4}\right).
\end{align}
From this result, we can also obtain the asymptotic form when $\overline{\Delta\mu}$ is large:
\begin{align}
V_{\mathcal{G}}
&\sim
C_{\delta_u}
\left(\overline{\Delta\mu}\right)^{-\delta_u},
\\
C_{\delta_u}
&:=
\sqrt{\pi}
\frac{
2^{\delta_u} 
\cos(\pi\delta_u)
}{
\Gamma\left(\frac{1-\delta_u}{2}\right)
\Gamma\left(\frac{2-\delta_u}{2}\right)
}.
\end{align}
We also note that the coefficient $C_{\delta_u}$ converges to 1 as $\delta_u\rightarrow 0$.

\subsection{Figures and discussions}
\label{subsec:calG:figs}
\begin{figure*}
	\centering
	\includegraphics[width=0.95\linewidth]{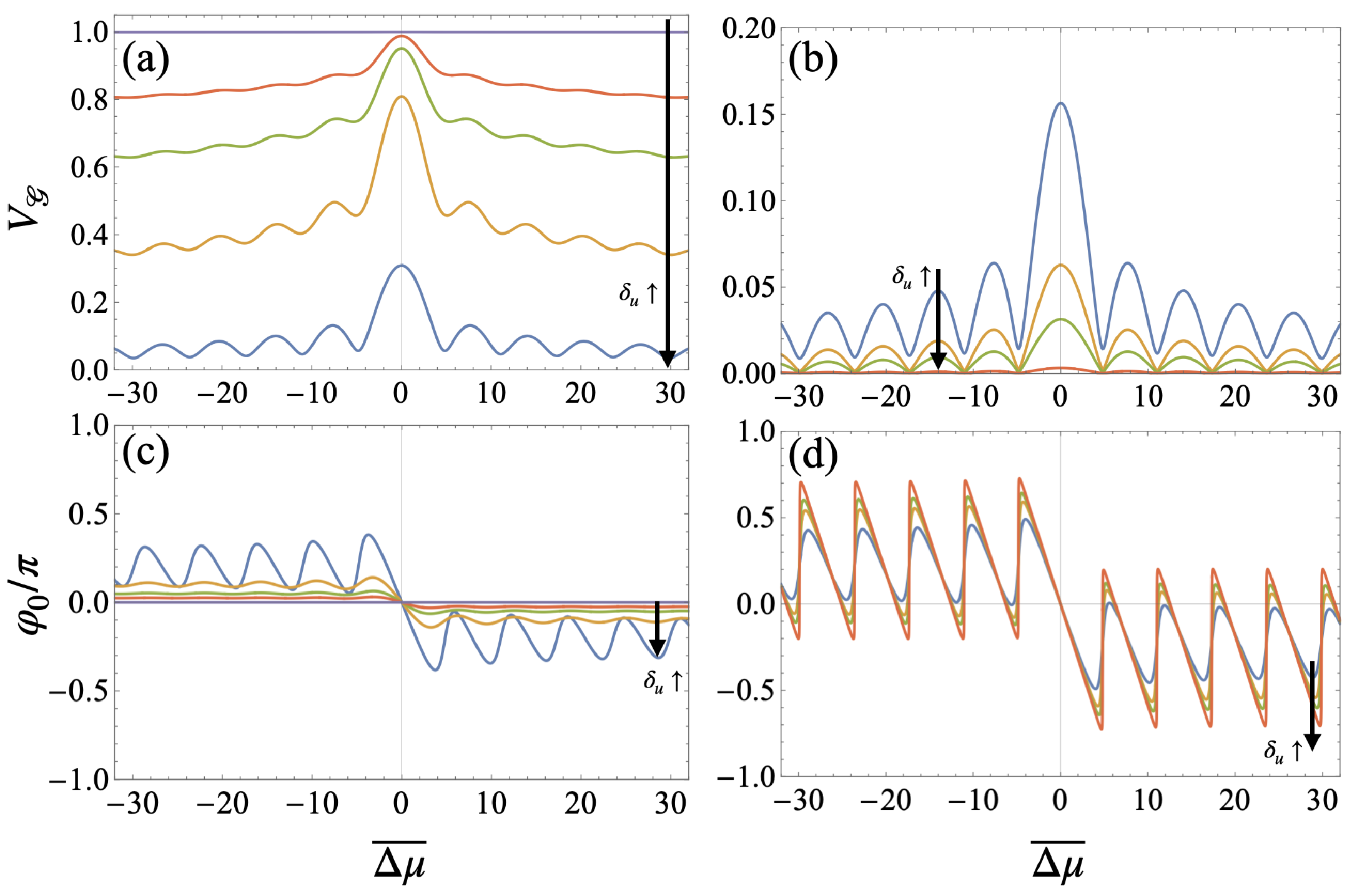}
	\caption{
	Bias dependence of $V_{\mathcal{G}}$ (a)(b) and the phase (c)(d) of AB oscillations at $|t_\rL|=|t_\rR|$.
	(a)(c): $\delta_u=0.0001,0.05,0.1,0.2,0.4$, 
    (b)(d): $\delta_u=0.499,~0.49,~0.48,~0.45$.
	}
	\label{fig:visibility}
\end{figure*}

Previous experiments on conventional MZIs fabricated in AlGaAs/GaAs heterostructures [Fig.~\ref{fig:MZI_setup}(a)] have shown oscillatory bias dependence of the visibility, which is often referred to as the lobe structure, accompanied by the non-monotonical phase evolution of the AB oscillations~\cite{Neder2006}. The phase jumps at the bottoms of the lobe structure and, if the interference path lengths are the same, the phase is kept constant at the other bias. Theories have successfully explained these observations by considering the impact of the inter-channel Coulomb interaction between the interference path and its environmental copropagating channel~\cite{Levkivskyi2008}.

Here, we numerically demonstrate the results for our setup obtained in Sec.~\ref{subsec:calG:analytic}. We observe an oscillatory visibility behavior similar to the conventional MZI but accompanied by different behaviors in the phase evolution. 

Figures~\ref{fig:visibility}(a) and (b) show the $\overline{\Delta\mu}$ dependence of the visibility $V_{\mathcal{G}}$. 
While we observe $V_{\mathcal{G}} = 1$ over the entire range of $\overline{\Delta\mu}$ at $\delta_u = 0$, $V_{\mathcal{G}}$ decreases with $|\overline{\Delta\mu}|$ at $\delta_u>0$, showing oscillatory behavior at finite $|\overline{\Delta\mu}|$. The visibility oscillations become more pronounced at larger values of $\delta_u$, indicating that the mode transformation affects the AB interference when the copropagating channels are electrostatically asymmetric. This observation can be interpreted as the result of the second-order interference between the fractionalized spin excitations.

Figures~\ref{fig:visibility}(c) and (d) show the phase $\varphi_0$ of the AB oscillations as a function of the bias. When $\delta_u > 0$, the phase jumps at the bottoms of the visibility oscillations, as observed in the conventional MZIs. Moreover, we observe linear bias dependence at the other $\overline{\Delta\mu}$ values, including $\overline{\Delta\mu}\simeq0$, even though our MZI has the same interference path lengths. The linear phase evolution is proportional to the asymmetry factor $\delta_u$ as
\begin{align}
\label{eq:ABphase_diff1}
\left.\frac{\partial\Delta\varphi}{\partial\Delta\mu}\right|_{\Delta\mu=0}=-\delta_u\frac{L}{v}.
\end{align}
This linear behavior reflects the difference between the speeds of the eigenmodes, instead of the different path lengths in the conventional setup.

\begin{figure}
	\centering
	\includegraphics[width=0.95\linewidth]{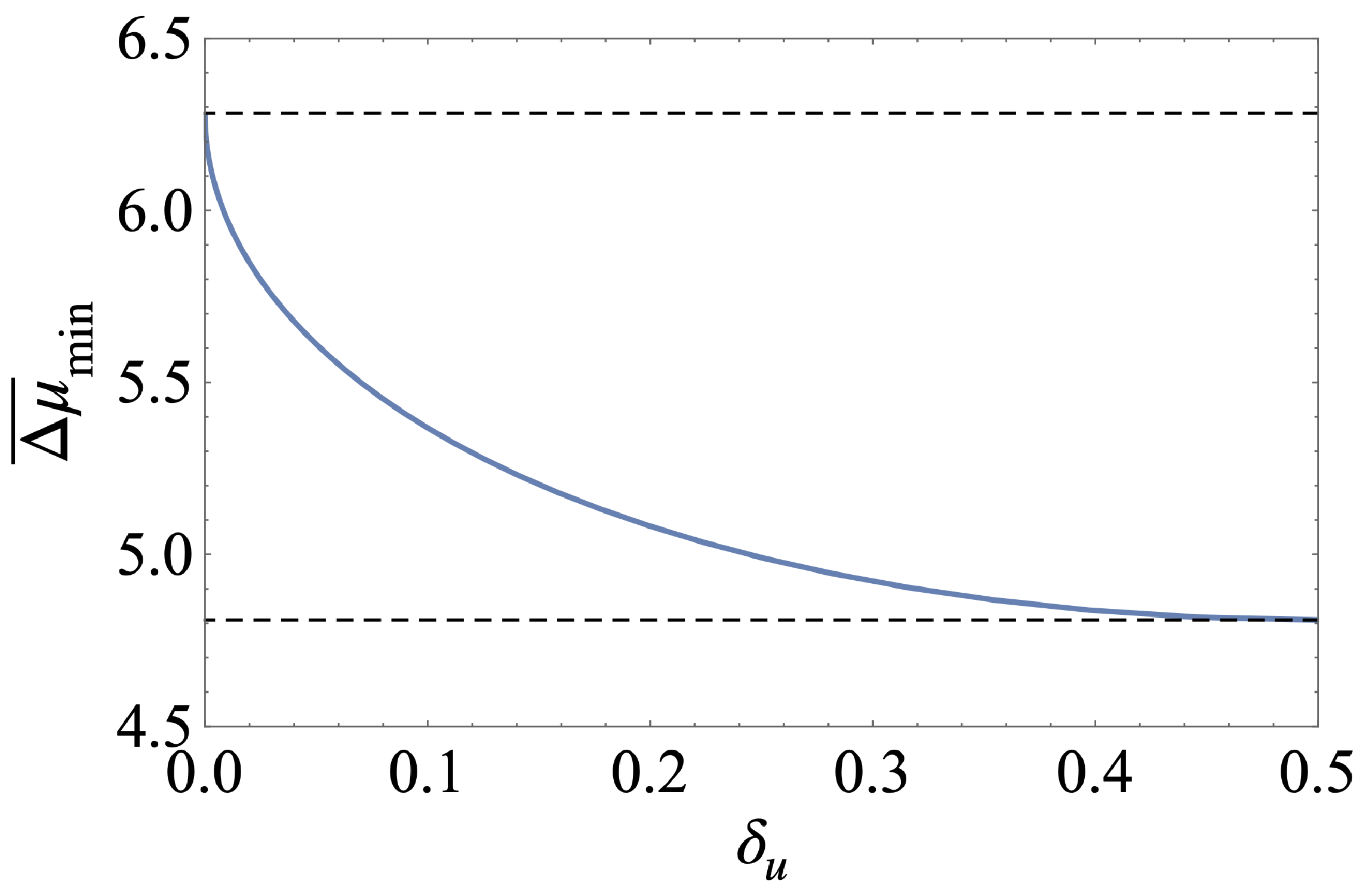}
	\caption{
	$\delta_u$ dependence of the first minimum of the lobe structure.
	The dashed lines represent $2\pi$ and $\overline{\Delta\mu}_\mathrm{min}(\delta_u=1/2)\simeq4.81$, respectively.
	}
	\label{fig:Lobe_1stmin}
\end{figure}

When $\delta_u=1/2$, the first minimum of the lobe pattern $\overline{\Delta\mu}_\mathrm{min}$ corresponds to the first zero of the Bessel function $J_0(\overline{\Delta\mu}/2)$; therefore, we find $\overline{\Delta\mu}_\mathrm{min}(\delta_u=1/2)\simeq 4.81$. $\overline{\Delta\mu}_\mathrm{min}$ monotonically increases toward $2\pi$ with decreasing $\delta_u$, as shown in Fig.~\ref{fig:Lobe_1stmin}. Since the normalized bias is defined as $\overline{\Delta\mu}:=L\Delta\mu/v$, 
the above result indicates that the oscillatory visibility behavior reflects the difference between the speeds and the resultant phase evolutions of the fast and slow modes.

\subsection{Remarks}
\label{subsec:calG:discussion}

The reason why the visibility shows the lobe pattern [see Fig.~\ref{fig:visibility}] is because there is the branch cut singularity in Eq.~(\ref{eq:current_2ch_ILR}), which results from the power-law decay of the correlation function. However, when the interaction is symmetric, the singularity becomes simply a pole, resulting in the absence of the visibility oscillations in Fig.~\ref{fig:visibility}.

While we have investigated an MZI employing two copropagating channels as the interference paths, we can extend our analysis to the other setups, e.g., by involving another environmental chiral channel coupled with the paths through Coulomb interaction. For instance, we present an analysis considering the third channel in Appendix~\ref{sec:setup:ThreeCh}.

For the above calculations, we have considered the lowest-order contribution with respect to the tunnel coupling [see Eq.~(\ref{eq:current_lowest_T})], assuming small transmission probabilities of the QPCs. The visibility oscillations resulting from the second-order interference of the fast/slow modes are expected to occur even at higher transmission probabilities. We discuss this point in Appendix~\ref{sec:appendix:current_scattering}. 

\section{Discussion}
\label{sec:discussion}

The above calculation results capture major features of the experimental results reported in our companion paper~\cite{Shimizu2024Joint}. In the experimental work, we measured an MZI device similar to the present setup shown in Fig.~\ref{fig:MZI_setup}(b). The visibility and the phase of the AB oscillations displayed in Fig.~\ref{fig:visibility} are qualitatively consistent with the experimental observations; particularly, we observe the visibility oscillations accompanied by the linear AB phase shift near zero bias both in the theory and the experiment. Thus, by considering only the inter-channel interaction that governs electron dynamics in the copropagating channels, we have successfully explained key features of the coherent evolution of a single electron under fractionalization. 

We note that the calculation and experimental results are not in complete quantitative agreement. The actual experimental device has various factors not considered in the present theoretical model, such as the inter-channel interaction with environmental channels, which we briefly discuss in Appendix~\ref{sec:setup:ThreeCh}. Moreover, our theory includes tunnel couplings and the asymmetric parameter perturbatively to simplify calculations. A more detailed examination of these factors would allow for more quantitative comparisons.

\section{Summary}
\label{sec:summary}

We have theoretically investigated single electron coherence under spin fractionalization in an artificial chiral TL liquid of copropagating QH edge channels. Previous studies investigated the fractionalization process by injecting a spin-up or a spin-down excitation into such copropagating channels. On the other hand, the present study analyzed the coherent evolution of a spin-up and spin-down superposition state by preparing an initial electronic state with a single scattering event at a QPC. We calculated electron transport through an MZI employing copropagating channels as the interference paths using the bosonization technique. While the single electron picture predicts the interference visibility independent of the bias for the one-way MZI, the electron splitting process can induce the visibility oscillations, the signature of the second-order interference between the fast and slow eigenmodes. The second-order interference occurs when the spin-up and spin-down channels are electrostatically asymmetric to cause the spin fractionalization. The spin fractionalization induces quantum entanglement between the fast and slow eigenmodes, resulting in their interference. The calculation results for the AB oscillations capture major features of the experimental results in our companion paper, suggesting the presence of quantum entanglement between the separated eigenmodes in the actual QH device. These findings highlight new functionalities of copropagating edge channels in electron quantum optics.

\begin{acknowledgments}
The authors thank S. Sasaki, A. Endo, S. Katsumoto, N. Kumada, and K. Muraki for fruitful discussions. This work was supported by JSPS KAKENHI Grant Nos. 19H05603, 22H00112, 23H01093, and 24H00827 and JSPS Bilateral Program Number JPJSBP120249911.
\end{acknowledgments}
\appendix

\section{Renyi-2 entanglement entropy}
\label{sec:appendixEE}
Here, we calculate the Renyi-2 EE by using the partial swap. According to Refs.~\cite{Hastings2010,Cardy2011,Herdman2014,Kaneko2020,Srdinsek2023}, the bipartition Renyi-2 EE of a state $|\psi\rangle$ is calculated as
\begin{align}
S_2(\rX):=
-\log
\left(
\langle\psi|\langle\psi|
C_\rX
|\psi\rangle|\psi\rangle
\right).
\end{align}
The operator $C_\rX$ is a partial swap operator and acts on a state and its copy, swapping the degrees of freedom of its subsystem X.
When the states of X and Y are labeled by $x_i$ and $y_i$, respectively, $C_\rS$ acts on elements of $(\rX\otimes \rY)^{\otimes 2}$ as follows:
\begin{align}
C_\rX|x_1 y_1\rangle|x_2 y_2\rangle
=
|x_2y_1\rangle|x_1y_2\rangle.
\end{align}

Let $\rX$ be the Fock space of the fast mode, and $\rY$ that of the slow mode in the present case.
By calculating $\langle\psi(t)|\langle\psi(t)|
C_\rX
|\psi(t)\rangle|\psi(t)\rangle$ paying attention to the Klein factor,
we obtain
\begin{align}
T^2+R^2+2TR 
\langle e^{i\phi_{\rU u}}e^{-i\phi_{\rL u}}\rangle
\langle e^{i\phi_{\rL u}}e^{-i\phi_{\rU u}}\rangle,
\end{align}
which leads to Eq.~(\ref{eq:Renyi-2EE}).
We note that correlations between the Klein factors of the upper and lower channels, e.g., $\langle F_\rU F_\rL^\dag\rangle$, are zero due to the electron number mismatch.

\section{Calculation of correlation function}
\label{sec:appendix2ch}

Here, we present the derivation of Eq.~(\ref{eq:result2ch4pt}). First, we show that some commutators of the bosonic fields are c-number (e.g., $[\phi_{\rU j},\phi_{\rL j}]\in\mathbb{C}$.) This fact validates the usage of the special cases of the Baker-Campbell-Hausdorff formula $e^A e^B=e^{A+B}e^{\frac{1}{2}[A,B]}$, which holds when $[A,B]\in\mathbb{C}$, to obtain Eq.~(\ref{eq:com_without_phase2}). 

We write $\phi_{ij}$ ($i=\rU,\rL$ and $j=u,v$) in Eq.~(\ref{eq:com_without_phase}) as
\begin{align}
\phi_{ij}
&=
\sum_{k>0}
\sqrt{
\frac{2\pi}{Wk}
}
\left[
S_{ij}b_j(k)e^{ikX_j}
+
S^*_{ij}b^\dag_j(k)e^{-ikX_j}
\right],
\end{align}
where $X_u
:=
x-ut$ and
$X_v
:=
x-vt$.
The unitarity of the S matrix leads to
\begin{align}
[\phi_{\rU j},\phi_{\rL j}]
&=
\sum_{k>0}
\frac{2\pi}{Wk}
\left(
S_{\rU j}(S_{\rL j})^*
-
S_{\rL j}(S_{\rU j})^*
\right).
\label{eq:cor_phi_UD}
\end{align}
This equation means $[\phi_{\rU j},\phi_{\rL j}]\in\mathbb{C}$, and we find
\begin{align}
\prod_{j=u,v}
e^{\frac{1}{2}[\phi_{\rU j},\phi_{\rL j}]}
=1
\end{align}
from simple algebra, using the unitarity again.
Note that the right hand side of Eq.~(\ref{eq:cor_phi_UD}) equals zero when the $S$ matrix is real. In the same way, we can also confirm $[\phi^\prime_{\rL j},\phi^\prime_{\rU j}]\in\mathbb{C}$.

We next calculate $[\phi_{\rU j}-\phi_{\rL j}, \phi^\prime_{\rU j}-\phi^\prime_{\rL j}]$.
Using the commutation relations between the boson fields, we obtain
\begin{align}
[
\phi_{\rU j}-\phi_{\rL j}
,
\phi^\prime_{\rU j}-\phi^\prime_{\rL j}
]
=
2i
\delta_j
\sum_{k>0}
\frac{2\pi}{W}\frac{\sin(k(X_j-X_j^\prime))}{k}.
\end{align}
The summation on the right-hand side can be replaced with the integral and calculated as
\begin{align}
\sum_{k>0}\frac{2\pi}{W}\frac{\sin(k(X_j-X_j^\prime))}{k}
&=
\int_0^\Lambda
dk
\frac{\sin(k(X_j-X_j^\prime))}{k}
\\
&\simeq \frac{\pi}{2}\sgn(X_j-X_j^\prime),
\end{align}
where $\Lambda$ is the ultraviolet cutoff.
Thus, we obtain
\begin{align}
[
\phi_{\rU j}-\phi_{\rL j}
,
\phi^\prime_{\rU j}-\phi^\prime_{\rL j}
]
&\simeq
\pi i
\delta_j
\sgn(X_j-X_j^\prime).
\end{align}

From the above calculations, the four-point function Eq.~(\ref{eq:com_without_phase}) is described as
\begin{align}
&
\mathcal{F}_{ll^\prime}(\delta,t)
\prod_{j=u,v}
\langle e^{-i C_j}\rangle,
\end{align}
except for the phase factor. Here, we define
\begin{align}
C_j:=&
\phi_{\rU j}-\phi_{\rL j}+\phi^\prime_{\rL j}-\phi^\prime_{\rU j}
=\delta\phi_{\rU j}-\delta\phi_{\rL j}
\\
\delta \phi_{ij}
:=&
\phi_{ij}-\phi_{ij}^\prime,
\\
\mathcal{F}_{ll^\prime}(\delta,t):=&
(-1)\times
\begin{cases}
\exp(\pi i \delta)& l\neq l^\prime\land \frac{L}{u}<t<\frac{L}{v}\\
1&\mathrm{otherwise}.
\end{cases}
\end{align}
We note that $\mathcal{F}_{ll^\prime}(\delta=0,t)=-1$ for any $t,l,$ and $l^\prime$ when the interaction matrix is completely symmetric [see Eq.~(\ref{eq:specialV})].

Since the fast and slow modes are eigenmodes of the present system, we can use the Gaussian character $\langle e^{-i C_j}\rangle=
e^{\frac{1}{2}\langle (-iC_j)^2\rangle}$ (Ref.~\cite{Shimizu2024Joint}) to obtain
\begin{align}
\langle e^{-i C_j}\rangle
&=
e^{\frac{1}{2}\langle (-iC_j)^2\rangle}
=
\exp
\left[
-\frac{1}{2}\sum_{i,i^\prime}\sigma_i \sigma_{i^\prime}
\langle
\delta \phi_{ij}
\delta \phi_{i^\prime j}
\rangle
\right],
\end{align}
where $\sigma_\rU=1$ and $\sigma_\rL=-1$.
At zero temperature limit, 
\begin{align}
\langle
\delta \phi_{ij}
\delta \phi_{i^\prime j}
\rangle
&\xrightarrow{T\rightarrow 0}
\sum_{k>0}
\frac{4\pi}{Wk}
S_{ij}(S_{i^\prime j})^*
\left(1-\cos(k(X_j-X_j^\prime))\right)
\\
&=
2S_{ij}(S_{i^\prime j})^*
\int_0^\Lambda dk \frac{1-\cos(k(X_j-X_j^\prime))}{k}
\\
&\simeq
S_{ij}(S_{i^\prime j})^*
\log\left(1+\frac{(X_j-X_j^\prime)^2}{\alpha^2}\right),
\end{align}
where we denote $\Lambda^{-1}$ as $\alpha$.
Thus, we find
\begin{align}
\langle e^{-iC_j}\rangle
&=
\frac{\alpha^{\delta_j}}{
\left(
(X_j-X_j^\prime)^2+\alpha^2
\right)^{\delta_j/2}
}.
\end{align}
and, consequently, Eq.~(\ref{eq:result2ch4pt}).

\section{Three-channel case}
\label{sec:setup:ThreeCh}

\begin{figure}[t]
	\centering
	\includegraphics[width=0.6\linewidth]{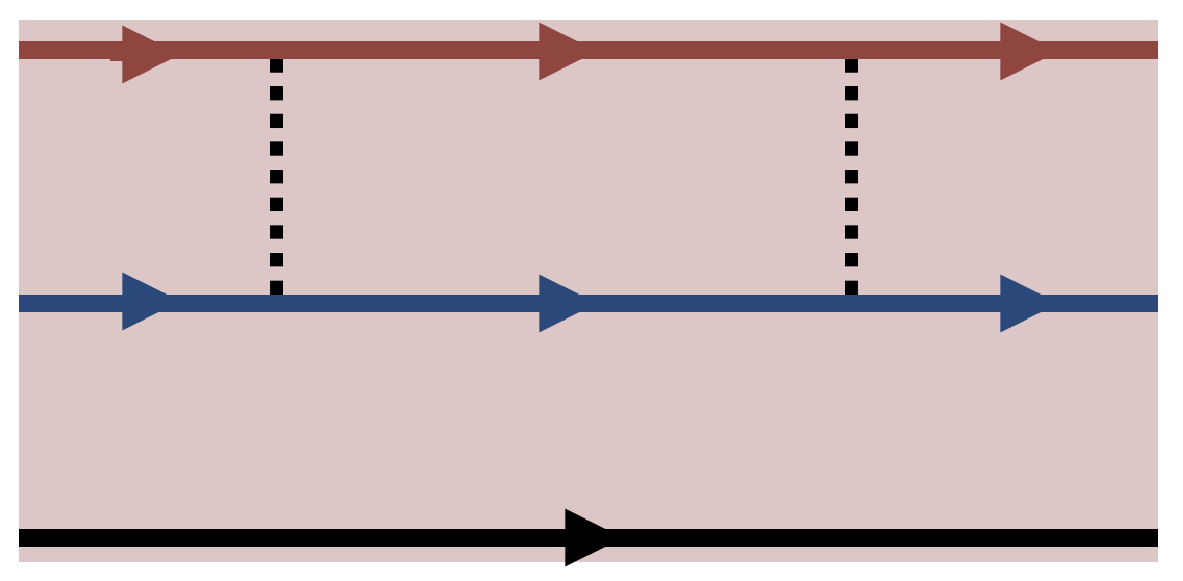}
	\caption{
		Schematic of MZI with an environmental channel. Red and blue lines are the main channels, and black one is environmental channel. Red region indicates the presence of Coulomb interaction.
	}
	\label{fig:system_three_channel}
\end{figure}
\begin{figure}[t]
	\centering
	\includegraphics[width=0.8\linewidth]{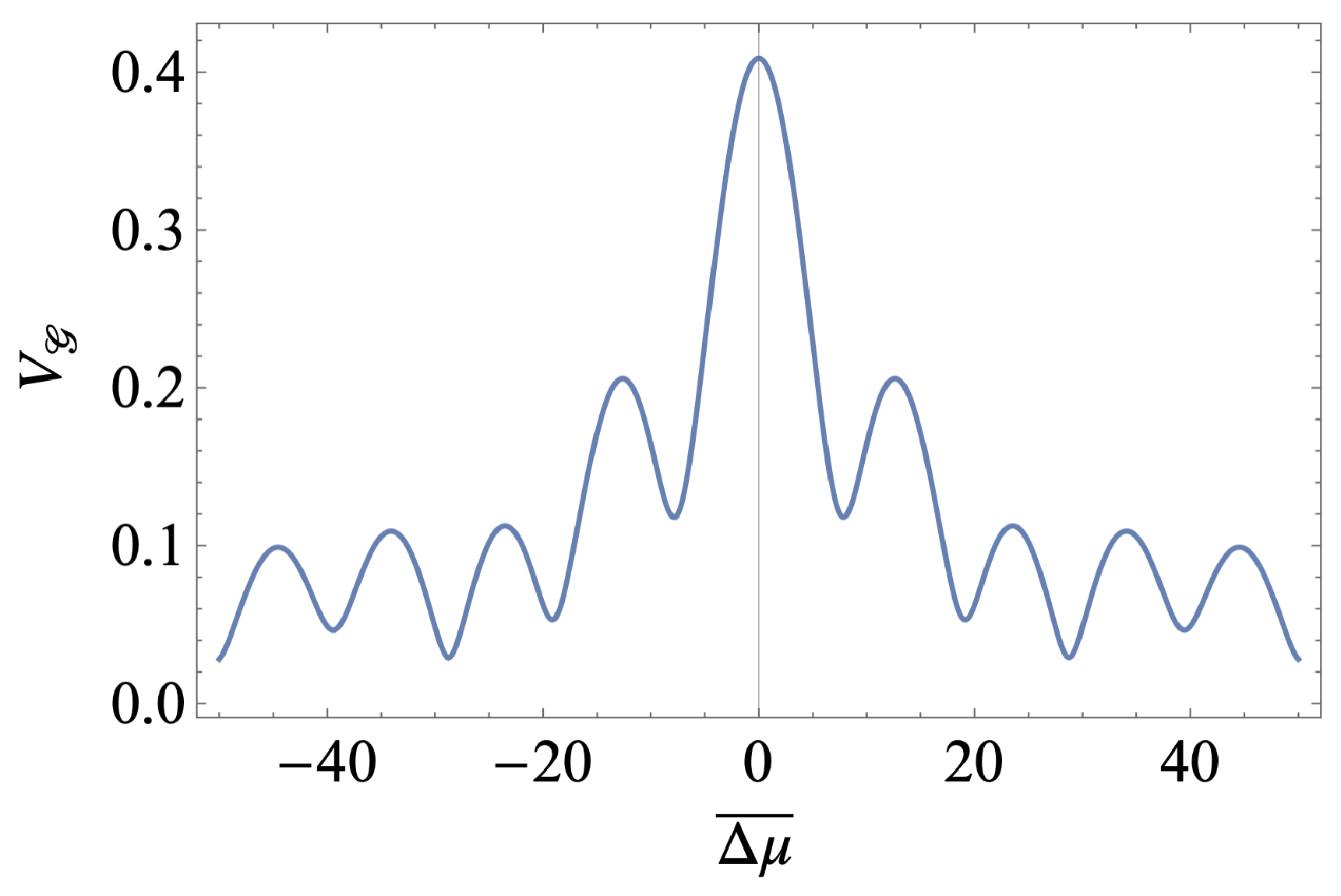}
	\caption{
	The bias dependence of the visibility in the three-channel case.
	We chose the parameters $u\simeq 305,~w\simeq 145,~v\simeq 58,~\delta_u\simeq 0.12,~\delta_w\simeq 1.47,$ and $\delta_v\simeq 0.41$, which could be close to those in the actual device reported in the companion paper.
	}
	\label{fig:visibility3ch}
\end{figure}

Here, we briefly investigate the AB oscillations when the third environmental channel copropagate the interference paths of the MZI, as shown in Fig.~\ref{fig:system_three_channel}. Note that the third channel couples with the main channels only through Coulomb interaction. The AB oscillations in this case can be calculated by essentially the same methodology as the two-channel case discussed in the main text. The interaction matrix is given by
\begin{align}
\label{eq:interaction_3ch1}
V
&=
\left[
 \begin{matrix}
   V_{\rU \rU} & V_{\rU \rL} & V_{\rU \rA} \\
   V_{\rL \rU} & V_{\rL \rL} & V_{\rL \rA}  \\
   V_{\rA \rU} & V_{\rA \rL} & V_{\rA \rA}  
 \end{matrix}
\right]
=
S\Lambda S^\dag,
\\
\label{eq:interaction_3ch2}
\Lambda&=
2\pi
\left[
 \begin{matrix}
   u & 0 & 0\\
   0 & v & 0 \\
   0 & 0 & w
 \end{matrix}
\right].
\end{align}

The average current can be mostly calculated in the same way as the two-channel case. The integral in Eq.~(\ref{eq:current_lesser}) is modified as
\begin{align}
\label{eq:current_lesser3ch}
\int dt 
\mathcal{F}_{ll^\prime}(\delta,t)
\frac{\gamma e^{i\Delta \mu (t-\Delta t_{ll^\prime})}}{
(X_u-X_u^\prime)^{\delta_u}
(X_w-X_w^\prime)^{\delta_w}
(X_v-X_v^\prime)^{\delta_v}
},
\end{align}
where the exponents are defined as
\begin{align}
\delta_j
&:=
1-|S_{\rA j}|^2-S_{\rU j} (S_{\rL j})^*-S_{\rL j} (S_{\rU j})^*~~~(j=u,w,v).
\end{align}
When $\delta_u\ll\delta_w\ll\delta_v$, the contribution from the branch cut between 
$X_u-X_u^\prime=0$ and $X_w-X_w^\prime=0$ is negligible. Consequently, the average current is obtained by replacing $\delta_u$ with $\delta_w$ and $u$ with $w$ in Eq.~(\ref{eq:current_lesser}) and performing the same calculations as the two channel case. Thus, the first minimum of the visibility oscillations is determined by the slow-mode speed.

Figure \ref{fig:visibility3ch} shows a numerically simulated AB-oscillation visibility in a general case, where $\delta_u\ll\delta_w\ll\delta_v$ does not hold. In such a case, the lobe pattern is slightly distorted due to the coupling with the third channel.

\section{AB oscillations at high transmission probabilities}
\label{sec:appendix:current_scattering}

We define the output current as the difference of the number of electrons between the upper and lower channels at the exit of the MZI:
\begin{align}
I\equiv
\left.
(
\psi_\rL^\dag \psi_\rL
-
\psi_\rU^\dag \psi_\rU)
\right|_{x=x_\rR+\varepsilon}
\end{align}
According to the scattering approach [Eq.~(\ref{eq:scattering_matrix})], the current can be written as
\begin{align}
\nonumber
I=&
(-i\sqrt{R}\psi_\rU^\dag +\sqrt{T}\psi_\rL^\dag)
(i\sqrt{R}\psi_\rU+\sqrt{T}\psi_\rL)
\\
&-
\left.
(\sqrt{T}\psi_\rU^\dag-i\sqrt{R}\psi_\rL^\dag)
(\sqrt{T}\psi_\rU+i\sqrt{R}\psi_\rL)\right|_{x=x_\rR-\varepsilon},
\end{align}
by pulling back operators at $x_\rR+\varepsilon$ to $x_\rR-\varepsilon$.

We here define $\mathcal{I}(t)$ as the bias-dependent portion of the expectation value $\langle\psi(t)|I|\psi(t)\rangle$ and focus on the integration of $\mathcal{I}(t)$ in $L/u\leq t\leq L/v$.
From simple algebra, we obtain
\begin{align}
\int_{L/u}^{L/v}
dt
\mathcal{I}(t)
&=
\frac{4RT}{u^{\delta_u}v^{\delta_v}}
\int_0^1
d\tau
\frac{
\cos(\overline{\Delta\mu}(\tau-1)-\varphi)
}{
\tau^{\delta_u}(1-\tau)^{\delta_v}
}.
\end{align}
Maximizing this with respect to the AB phase $\varphi$, we find
\begin{align}
\max_\varphi
\int_{L/u}^{L/v}
dt
\mathcal{I}(t)
\propto
4RTV_{\mathcal{G}}.
\end{align}
The right hand side of this equation contains the visibility [Eq.~(\ref{eq:visibility1})], which is perturbatively derived with respect to the tunneling term $\mathcal{H}_\rT$.
This result implies that the visibility oscillations also appear at high QPC transmission probabilities.


\begin{thebibliography}{99}
\bibitem{Andreev1963}
A.F. Andreev,
{\it Andreev reflection},
{\it Sov. Phys. JETP} {\bf 19}, 1228 (1964).

\bibitem{Tinkham1996}
M. Tinkham, 
{\it textbook about Superconductivity}
Introduction to Superconductivity (McGraw Hill, 1996).

\bibitem{Hashisaka2021}
M. Hashisaka, T. Jonckheere, T. Akiho, S. Sasaki, J. Rech, T. Martin, and K. Muraki,
{\it FQH Andreev-like reflection}
{\it Nat. Commun.} {\bf 12}, 2794 (2021).

\bibitem{Cohen2023}
L.A. Cohen, N.L. Samuelson, T. Wang, T. Taniguchi, K. Watanabe, M.P. Zaletel, and A.F. Young,
{\it FQH Andreev-like reflection}
{\it Science} {\bf 382}, 542 (2023).

\bibitem{Tomonaga1950}
S. Tomonaga,
{\it Remarks on Bloch’s Method of Sound Waves Applied to Many-Fermion Problems},
{\it Prog. Theor. Phys.} {\bf 5}, 544 (1950).

\bibitem{Luttinger1963}
J.M. Luttinger, {\it An Exactly Soluble Model of a Many-Fermion System},
{\it J. Math. Phys.} {\bf 4}, 1154 (1963).

\bibitem{Haldane1981}
F.D.M. Haldane, {\it `Luttinger Liquid Theory' of One-Dimensional Quantum Fluids. I. Properties of the Luttinger Model and Their Extension to the General 1D Interacting Spinless Fermi Gas}, {\it J. Phys. C} {\bf 14}, 2585 (1981).

\bibitem{Giamarchi2003}
T. Giamarchi, 
{\it Quantum Physics in One Dimension}
(University Press, Oxford, 2003).

\bibitem{Chang2003}
A. M. Chang,
{\it Chiral Luttinger Liquids at the Fractional Quantum Hall Edge},
{\it Rev. Mod. Phys.} {\bf 75}, 1449 (2003).

\bibitem{Auslaender2002}
O. M. Auslaender, A. Yacoby, R. de Picciotto, K. W. Baldwin, L. N. Pfeiffer, and K. W. West,
{\it Tunneling Spectroscopy of the Elementary Excitations in a One-Dimensional Wire},
{\it Science} {\bf 295}, 825 (2002).

\bibitem{Jompol2009}
Y. Jompol, C. J. B. Ford, J. P. Griffiths, I. Farrer, G. A. C. Jones, D. Anderson, D. A. Ritchie, T. W. Silk, and A. J. Schofield,
{\it Probing Spin-Charge Separation in a Tomonaga–Luttinger liquid},
{\it Science} {\bf 325}, 597 (2009).

\bibitem{Berg2009}
E. Berg, Y. Oreg, E.-A. Kim, and F. von Oppen, 
{\it Fractional Charges on an Integer Quantum Hall Edge},
{\it Phys. Rev. Lett.} {\bf 102}, 236402 (2009).

\bibitem{Bocquillon2013}
E. Bocquillon, V. Freulon, P. Degiovanni, B. Plaçais, A. Cavanna, Y. Jin, and G. Fève, 
{\it Separation of neutral and charge modes in one-dimensional chiral edge channels},
{\it Nat. Commun.} {\bf 4}, 1839 (2013).

\bibitem{Kamata2014}
H. Kamata, N. Kumada, M. Hashisaka, K. Muraki, and T. Fujisawa,
{\it Fractionalized Wave Packets from an Artificial Tomonaga–Luttinger Liquid},
{\it Nat. Nanotech.} {\bf 9}, 177 (2014).

\bibitem{Inoue2014}
H. Inoue, A. Grivnin, N. Ofek, I. Neder, M. Heiblum, V. Umansky, and D. Mahalu, 
{\it Charge Fractionalization in the Integer Quantum Hall Effect},
{\it Phys. Rev. Lett.} {\bf 112}, 166801 (2014).

\bibitem{Freulon2015}
V. Freulon, A. Marguerite, J.-M. Berroir, B. Placais, A. Cavanna, Y. Jin, and G. Feve,
{\it Hong–Ou–Mandel experiment for temporal investigation of single-electron fractionalization},
{\it Nat. Commun.} {\bf 6}, 6854 (2015).

\bibitem{Hashisaka2017}
M. Hashisaka, N. Hiyama, T. Akiho, K. Muraki, and T. Fujisawa, 
{\it Waveform measurement of charge- and spin-density wavepackets in a chiral Tomonaga-Luttinger liquid},
{\it Nat. Phys.} {\bf 13}, 559 (2017).

\bibitem{Hashisaka2018}
M. Hashisaka and T. Fujisawa.
{\it Tomonaga–Luttinger-liquid nature of edge excitations in integer quantum Hall edge channels},
{\it Rev. in Phys.} {\bf 3}, 32 (2018).

\bibitem{Nakajima2013}
T. Nakajima, K.T. Lin, and S. Komiyama,
{\it copropagating MZI},
{\it AIP Conf. Proc.} {\bf 1566}, 301 (2013).

\bibitem{Shimizu2020}
T. Shimizu, T. Nakamura, Y. Hashimoto, A. Endo and S. Katsumoto,
{\it Gate-controlled unitary operation on flying spin qubits in quantum Hall edge states},
{\it Phys. Rev. B} {\bf 102} 235302 (2020).

\bibitem{Shimizu2023}
T. Shimizu, J. Ohe, A. Endo, T. Nakamura, and S. Katsumoto,
{\it Half-Mirror for Electrons in Quantum Hall Copropagating Edge Channels in a Mach-Zehnder Interferometer},
{\it Phys. Rev. Appl.} {\bf 19} 034085 (2023).

\bibitem{Shimizu2024Joint}
T. Shimizu, E. Iyoda, S. Sasaki, A. Endo, S. Katsumoto, N. Kumada, and M. Hashisaka, submitted (the companion paper of this manuscript).

\bibitem{Ji2003}
Y. Ji, Y. Chung, D. Sprinzak, M. Heiblum, D. Mahalu, and H. Shtrikman,
{\it An electronic Mach-Zehnder interferometer},
{\it Nature} {\bf 422} 415 (2003).

\bibitem{Neder2006}
I. Neder, M. Heiblum, Y. Levinson, D. Mahalu, and V. Umansky,
{\it Unexpected Behavior in a Two-Path Electron Interferometer},
{\it Phys. Rev. Lett.} {\bf 96} 016804 (2006).

\bibitem{Roulleau2007}
P. Roulleau, F. Portier, D. C. Glattli, P. Roche, A. Cavanna, G. Faini, U. Gennser, and D. Mailly, {\it Finite bias visibility of the electronic Mach-Zehnder interferometer}, {\it Phys. Rev. B} {\bf 76}, 161309(R) (2007).

\bibitem{Litvin2008}
L. V. Litvin, A. Helzel, H.-P. Tranitz, W. Wegscheider, and C. Strunk, 
{\it Edge-channel interference controlled by Landau level filling},
{\it Phys. Rev. B} {\bf 78}, 075303 (2008).

\bibitem{Bieri2009}
E. Bieri, M. Weiss, O. Goktas, M. Hauser, C. Schonenberger, and S. Oberholzer, 
{\it Finite-bias visibility dependence in an electronic Mach-Zehnder interferometer},
{\it Phys. Rev. B} {\bf 79}, 245324 (2009).

\bibitem{Sukhorukov2007}
E. V. Sukhorukov and V. Cheianov,
{\it Resonant Dephasing in the Electronic Mach-Zehnder Interferometer},
{\it Phys. Rev. Lett.} {\bf 99} 156801 (2007).

\bibitem{Chalker2007}
J. T. Chalker, Y. Gefen, and M. Y. Veillette,
{\it Decoherence and interactions in an electronic Mach-Zehnder interferometer},
{\it Phys. Rev. B} {\bf 76} 085320 (2007).

\bibitem{Neder2008}
I. Neder and E. Ginossar,
{\it Behavior of Electronic Interferometers in the Nonlinear Regime},
{\it Phys. Rev. Lett.} {\bf 100} 196806 (2008).

\bibitem{Youn2008}
S.-C. Youn, H.-W. Lee, and H.-S. Sim,
{\it Nonequilibrium Dephasing in an Electronic Mach-Zehnder Interferometer},
{\it Phys. Rev. Lett.} {\bf 100} 196807 (2008).

\bibitem{Helzel2015}
A. Helzel, L. V. Litvin, I. P. Levkivskyi, E. V. Sukhorukov, W. Wegscheider, and C. Strunk,
{\it Counting statistics and dephasing transition in an electronic Mach-Zehnder interferometer},
{\it Phys. Rev. B} {\bf 91}, 245419 (2015).

\bibitem{Jo2022}
M. Jo, J.-Y. M. Lee, A. Assouline, P. Brasseur, K. Watanabe, T. Taniguchi, P. Roche, D. C. Glattli, N. Kumada, F. D. Parmentier, H.-S. Sim, and P. Roulleau, 
{\it Scaling behavior of electron decoherence in a graphene Mach-Zehnder interferometer},
{\it Nat. Commun.} {\bf 13}, 5473 (2022).

\bibitem{Levkivskyi2008}
I. P. Levkivskyi and E. V. Sukhorukov,
{\it Dephasing in the electronic Mach-Zehnder interferometer at filling factor $\nu=2$},
{\it Phys. Rev. B} {\bf 78} 045322 (2008).

\bibitem{Hofstetter2009}
L. Hofstetter, S. Csonka, J. Nyg{\aa}rd, and C. Sch\"{o}nenberger,
{\it Cooper-pair splitting},
{\it Nature} {\bf 461}, 960 (2009).

\bibitem{Delft1998}
J. von Delft and H. Schoeller,
{\it Bosonization for beginners} \text{-} {\it refermionization for experts},
{\it Ann. Phys.} {\bf 7}, 225 (1998).

\bibitem{Rebora2020}
G. Rebora, M. Acciai, D. Ferraro, and M. Sassetti,
{\it Collisional interferometry of levitons in quantum Hall edge channels at $\nu=2$},
{\it Phys. Rev. B} {\bf 101}, 245310 (2020).

\bibitem{Acciai2022}
M. Acciai, P. Roulleau, I. Taktak, D. C. Glattli, and J. Splettstoesser,
{\it Influence of channel mixing in fermionic Hong-Ou-Mandel experiments},
{\it Phys. Rev. B} {\bf 105}, 125415 (2022).

\bibitem{Beenakker2006}
C.W.J. Beenakker, {\it Electron-hole entanglement in the Fermi sea},
{\it Proc. International School of Physics "Enrico Fermi"} {\bf 162}, 307-347 (2006).

\bibitem{Zhang2024}
G. Zhang, C. Hong, T. Alkalay, V. Umansky, M. Heiblum, I. Gornyi, and Y. Gefen, {\it Measuring statistics-induced entanglement entropy with a Hong-Ou-Mandel interferometer},
{\it Nat. Commun.} {\bf 15}, 3428 (2024).

\bibitem{Levkivskyi2010}
I. P. Levkivskyi and E. V. Sukhorukov, 
{\it Noise-Induced Phase Transition in the Electronic Mach-Zehnder Interferometer},
{\it Phys. Rev. Lett.} {\bf 103}, 036801 (2009).

\bibitem{Hastings2010}
M. B. Hastings, I. Gonzalez, A. B. Kallin, and R. G. Melko, 
{\it Measuring Renyi Entanglement Entropy in Quantum Monte Carlo Simulations},
{\it Phys. Rev. Lett.} {\bf 104}, 157201 (2010).

\bibitem{Cardy2011}
J. Cardy, 
{\it Measuring Entanglement Using Quantum Quenches},
{\it Phys. Rev. Lett.} {\bf 106}, 150404 (2011).

\bibitem{Herdman2014}
C. M. Herdman, P.-N. Roy, R. G. Melko, and A. Del Maestro,
{\it Particle entanglement in continuum many-body systems via quantum Monte Carlo},
{\it Phys. Rev. B} {\bf 89}, 140501(R) (2014).

\bibitem{Kaneko2020}
K. Kaneko, E. Iyoda, and T. Sagawa,
{\it Characterizing complexity of many-body quantum dynamics by higher-order eigenstate thermalization},
{\it Phys. Rev. A} {\bf 101}, 042126 (2020).

\bibitem{Srdinsek2023}
M. Srdinsek, M. Casula, and R. Vuilleumier,
{\it Renyi entropy of a quantum anharmonic chain at nonzero temperature},
{\it Phys. Rev. B} {\bf 108}, 245121 (2023).

\end{thebibliography}
\end{document}